\documentclass[final]{siamltex}
\usepackage{epsfig, graphicx,subfigure}%
\usepackage{graphicx}
\usepackage{amsmath}
\usepackage{amsfonts,amssymb}

\def\ds{\displaystyle}

\newcommand{\bea}{\begin{eqnarray}}
\newcommand{\eea}{\end{eqnarray}}

\newcommand{\p}{\partial}
\newcommand{\beq}{\begin{eqnarray}}
\newcommand{\beqq}{\begin{eqnarray*}}
\newcommand{\eeq}{\end{eqnarray}}
\newcommand{\eeqq}{\end{eqnarray*}}

\newcommand{\x}{\mbox{\boldmath$x$}}
\newcommand{\y}{\mbox{\boldmath$y$}}

\newcommand{\rr}{\mbox{\boldmath$r$}}

\newcommand{\n}{\mbox{\boldmath$n$}}

\title{Effective motion of a virus trafficking inside a biological cell}

\author{ Thibault Lagache \thanks{Department of Biology, Ecole Normale Sup\'erieure, Paris, France,
({\tt lagache@biologie.ens.fr}).}\and David Holcman
\thanks{Department of Mathematics, Weizmann Institute of Science, Rehovot 76100, Israel and
Department of Biology and Mathematics, Ecole Normale Sup\'erieure,
Paris, France.}}

\begin{document}

\maketitle

\begin{abstract}

Virus trafficking is fundamental for infection success and plasmid
cytosolic trafficking is a key step of gene delivery. Based on the
main physical properties of the cellular transport machinery such as
microtubules, motor proteins, our goal here is to derive a
mathematical model to study cytoplasmic trafficking. Because
experimental results reveal that both active and passive movement
are necessary for a virus to reach the cell nucleus, by taking into
account the complex interactions of the virus with the microtubules,
we derive here an estimate of the mean time a virus reaches the
nucleus. In particular, we present a mathematical procedure in which
the complex viral movement, oscillating between pure diffusion and a
deterministic movement along microtubules, can be approximated by a
steady state stochastic equation with a constant effective drift. An
explicit expression for the drift amplitude is given as a function
of the real drift, the density of microtubules and other physical
parameters. The present approach can be used to model viral
trafficking inside the cytoplasm, which is a fundamental step of
viral infection, leading to viral replication and in some cases to
cell damage.
\end{abstract}



\pagestyle{myheadings} \thispagestyle{plain} \markboth{T. LAGACHE AND D. HOLCMAN}{COMPUTATION OF THE MEAN FIELD INDUCED BY
MICROTUBULES IN CELLS}

\section{Introduction}
Because cytosolic transport has been identified as a critical
barrier for synthetic gene delivery \cite{Wiethoff}, plasmids or
viral DNAs  delivery from the cell membrane to the nuclear pores has
attracted the attention of many biologists. The cell cytosol
contains many types of organelles, actin filaments, microtubules and
many others, so that to reach the nucleus, a viral DNA has to travel
through a crowded and risky environment. We are interested here in
studying the efficiency of the delivery process and we present a
mathematical model of virus trafficking inside the cell cytoplasm.
We model the viral movement as a Brownian motion. However, the
density of actin filaments and microtubules, inside the cell, can
hinder diffusion, as demonstrated experimentally \cite{Dauty}.
 In a crowded environment, we will model the virus as a
material point. This reduction is simplistic for several reasons:
actin filament network can trapped a diffusing object and beyond a
certain size, as observed experimentally, a DNA fragment cannot find
its way across the actin filaments \cite{Dauty}. Active directional
transport along microtubules or actin filaments seems then the only
way to deliver a plasmid to the nucleus. The active transport of the
virus involves in general motor proteins, such as Kinesin (to travel
in the direction of the cell membrane) or Dynein (to travel toward
the nucleus). Once a virus is attached to a Dynein protein, its
movement can be modeled as a deterministic drift toward the nucleus.

Recently, a macroscopic modeling has been developed to describe the
dynamics of adenovirus concentration inside the cell cytoplasm
\cite{Dinh}. This approach offers very interesting results about
the effect of microtubules, but neglects the complexity of the
geometry and cannot be used to describe the movement of a single
virus, which might be enough to cause cellular infection. Modeling a
virus trafficking imposes to use a stochastic description. We model
here the motion of a virus as that of a material point, so the
probability of its trapping by actin filaments or microtubules is
neglected. In the present approximation, the viral movement has two
main components: a Brownian one, which accounts for its free
movement, and a drift directed towards the centrosome or MTOC
(Microtubules Organization Center), an organelle located near the
nucleus. The magnitude of the drift along microtubules depends on
many parameters, such as the binding and unbinding rates and the
velocity of the motor proteins \cite{alberts}.

In the present approach, we present a method to approximate a time
dependent dynamics of virus trafficking  by an effective stochastic
equation with a radial steady state drift. The main difficulties we
have to overcome arise from the time dependent nature of the
trajectories which consists of intermittent epochs of drifts and
free diffusion. We propose to derive an explicit expression for the
steady state drift amplitude. In this approximation, the effective
drift will gather the mean properties of the cytoplasmic
organization such as the density of microtubules and its off binding
rate.

Our method to find the effective drift can be described as follow:
first, we approximate the cell geometry as a two dimensional disk
and use a pure Brownian description to approximate the virus
diffusion step. This geometrical approximation is valid, for any two
dimensional cell  such as the
\textit{in vitro} flat skin fibroblast culture cells \cite{Dinh}: indeed, due to
their adhesion to the substrate, the thickness of these cells can be
neglected in first approximation. Second, when the distribution of
the initial viral position is uniform on the cell surface, we will
estimate, during the diffusing period, the hitting position on a
microtubule. By solving a partial differential equation, inside a
sliced shape domain, delimited by two neighboring microtubules, we
will provide an estimate of the mean time to the most likely hitting
point. Finally, the amplitude of the radial steady state drift will
be obtained by an iterative method which assumes that, after a virus
has moved a certain distance along a microtubule, it is released at
a point uniformly distributed on the final radial distance from the
nucleus, ready for a new random walk. This scenario repeats until
the virus reaches the nucleus surface. Finally, we will compute the
mean time, the mean number of steps before a virus reaches the
nucleus and the amplitude of the effective drift by using the
following criteria: the Mean First Passage Time (MFPT) to the
nucleus of the iterative approximation is equal to the MFPT obtained
by solving directly an Ornstein-Uhlenbeck stochastic equation. The
explicit computation of the effective drift is a key result in the
estimation of the probability and the mean time a single virus or
DNA molecule takes to reach a small nuclear pore \cite{david}.

\section{ Modeling stochastic viral movement inside a biological cell}
\begin{figure}[ht]
 \mbox{ \subfigure[]{\epsfig{figure=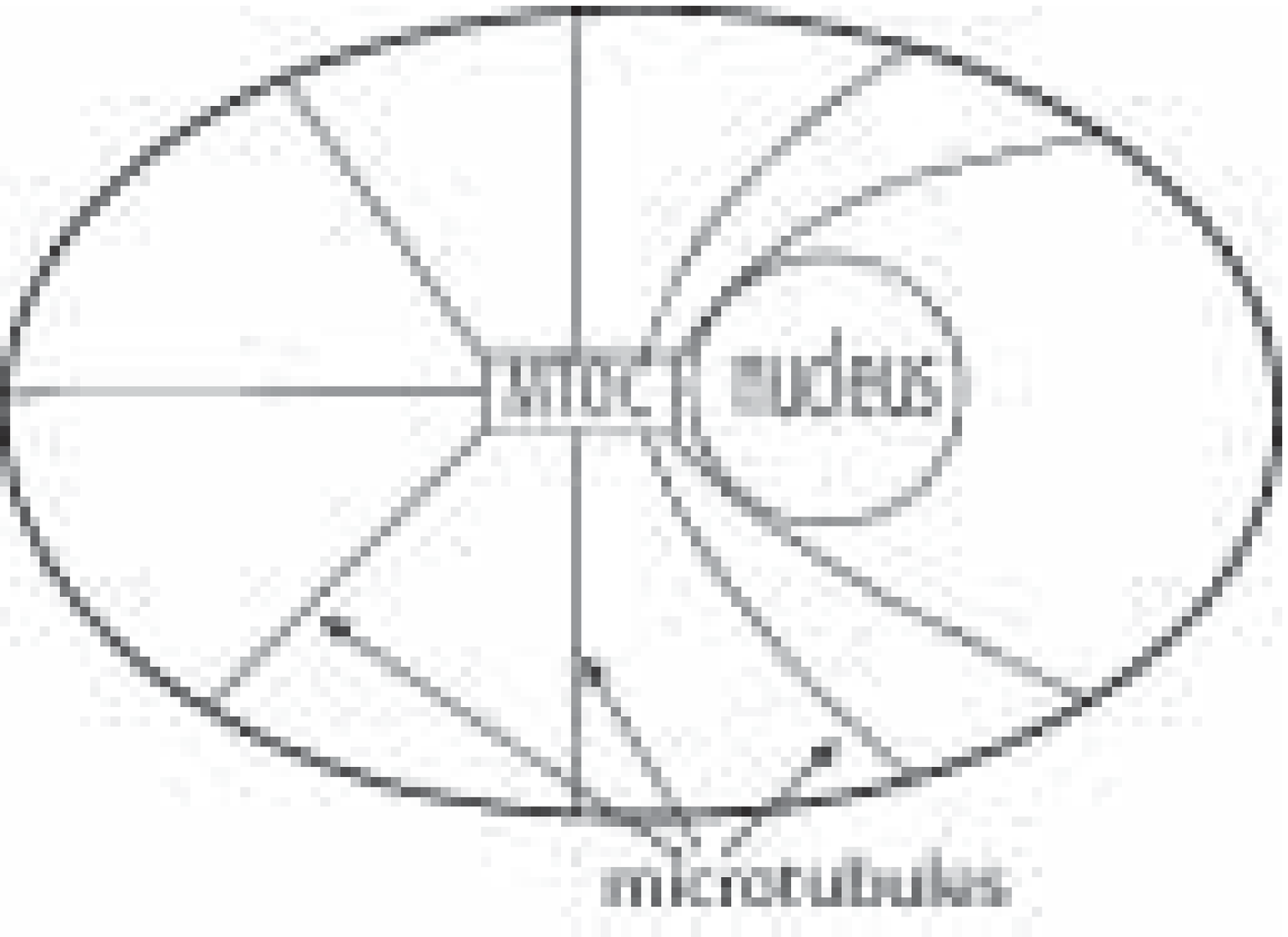,width=0.5\textwidth}}} \quad
 \mbox{ \subfigure[]{\epsfig{figure=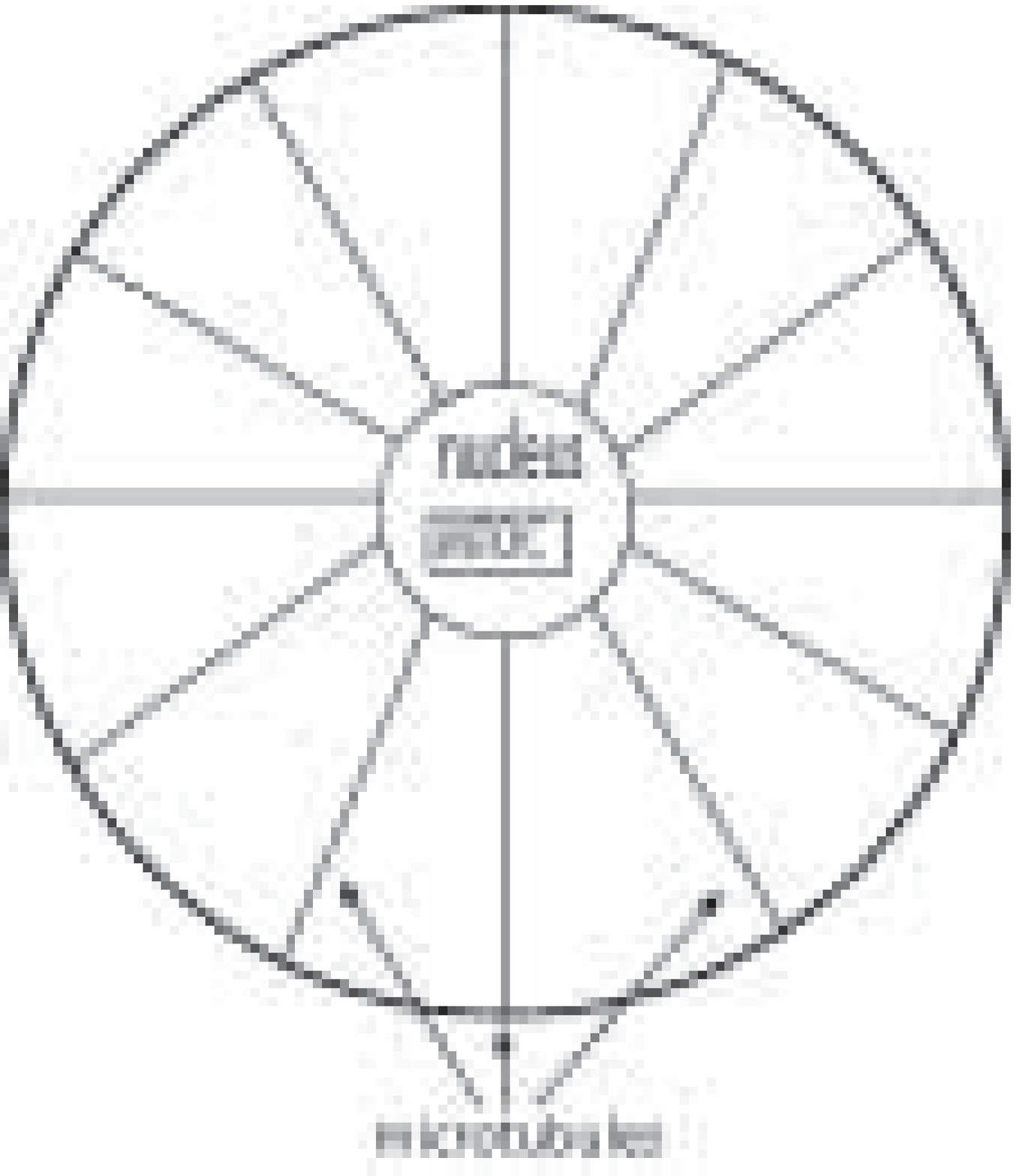,width=0.5\textwidth}}} \quad
\caption{{\bf Cell geometry.} (a) Cell's microtubules network. All microtubules starting from the cell membrane
converge to the Microtubule Organization center (MTOC), located near
the nucleus. (b) simplified cell's microtubules network
organization. The MTOC coincides with the nucleus.}
\label{network}
\end{figure}

We approximate the cell as a two dimensional geometrical domain
$\Omega $, which is here a disk of radius R and the nucleus located
inside is a concentric disk of much smaller radius $\delta<<R$. We
model the motion of an unattached DNA fragment as a material point,
so that the probability of its trapping by actin filaments or
microtubules is neglected. The motion of a (DNA) molecule of mass
$m$ is described by the overdamped limit of the Langevin equation
(Smoluchowski's limit) \cite{book} for the position  $\mathbf{X}(t)$
of the molecule at time $t$. When the particle is not bound to a
microtubule filament, its movement is described as pure Brownian
{with a diffusion constant $D$}. When the particle hits a filament,
it binds for a certain random time and moves along with a
determinist drift. We only take into account the movement toward the
nucleus, which is confound here with the MTOC (Microtubule
organization center), an organelle where all microtubules converge
(see figure (\ref{network})). For $\delta<|\mathbf{X}(t)|<R$, we
describe the overall movement by the stochastic rule
\bea \mathbf{\dot {X}}=
\left\{\begin{array}{l}  \sqrt {2D }
\mathbf{\dot {w}} \quad\mbox{ for }\quad  \mathbf{X}\left(t\right) \quad\mbox{ free }\\ \\
V \frac{\mathbf{r}}{|\mathbf{r}|} \quad\mbox{ for }\quad  \mathbf{X}\left(t\right)
\quad\mbox{ bound }
\end{array}\right.\label{eq1}
\eea
where $V$ is a constant velocity, $\mathbf{\dot {w}}$ a
$\delta$-correlated standard white noise and $\mathbf{r}$ the $\mathbf{X}$ radial
coordinate, the origin of which is the center of the cell. We
assume that all filaments starting from the cell surface end on the
nucleus surface. The binding time corresponds to a chemical reaction
event and we assume that it is exponentially distributed and for
simplicity we approximate it by a constant $t_m$.

Once a virus enters the cell membrane, its moves according to the
rule (\ref{eq1}), until it hits a nuclear pore. Although nuclear
pores occupy a small portion of the nuclear surface, we only
consider the virus movement until it hits the nuclear surface
$D\left(\delta\right)$.
%
In this article, our goal is to replace equation (\ref{eq1}) by a
steady state stochastic equation
 \bea \label{eq2} \dot
{\bf X}=\mathbf{b}(\mathbf{X}) + \sqrt {2 D } \mathbf{\dot {w}},
\eea
where the drift $\mathbf{b}$ is radially symmetric. In a first
approximation, we consider a constant radial drift
$\mathbf{b}(\mathbf{X})=-B
\frac{\mathbf{r}}{|\mathbf{r}|}$ and compute hereafter the value
of the constant amplitude $B$  such that the MFPT of the process
(\ref{eq2}) and (\ref{eq1}) to the nucleus are equal.
%
\subsection{ Modeling viral dynamics in the cytoplasm}
Inside the cytosol, microtubules are distributed on the cell surface
and  converging radially to the MTOC. We denote by $\rho$ this
distribution (see figure (\ref{network})). We do not take into
account in the present analysis, the effect of organelle crowding
due to the endoplasmic reticulum, the Golgi apparatus and many
others. However, it is always possible to include them indirectly by
using an apparent diffusion constant. We consider the fundamental
domain $\tilde{\Omega}$ defined as the two dimensional slice of
angle $\Theta$ between two neighboring microtubules. We consider
here that microtubules are uniformly distributed and thus $\Theta
=\frac{2\pi}{N}$, where $N$ is the total number of microtubules.

\begin{figure}[ht]
 \mbox{ \subfigure[]{\epsfig{figure=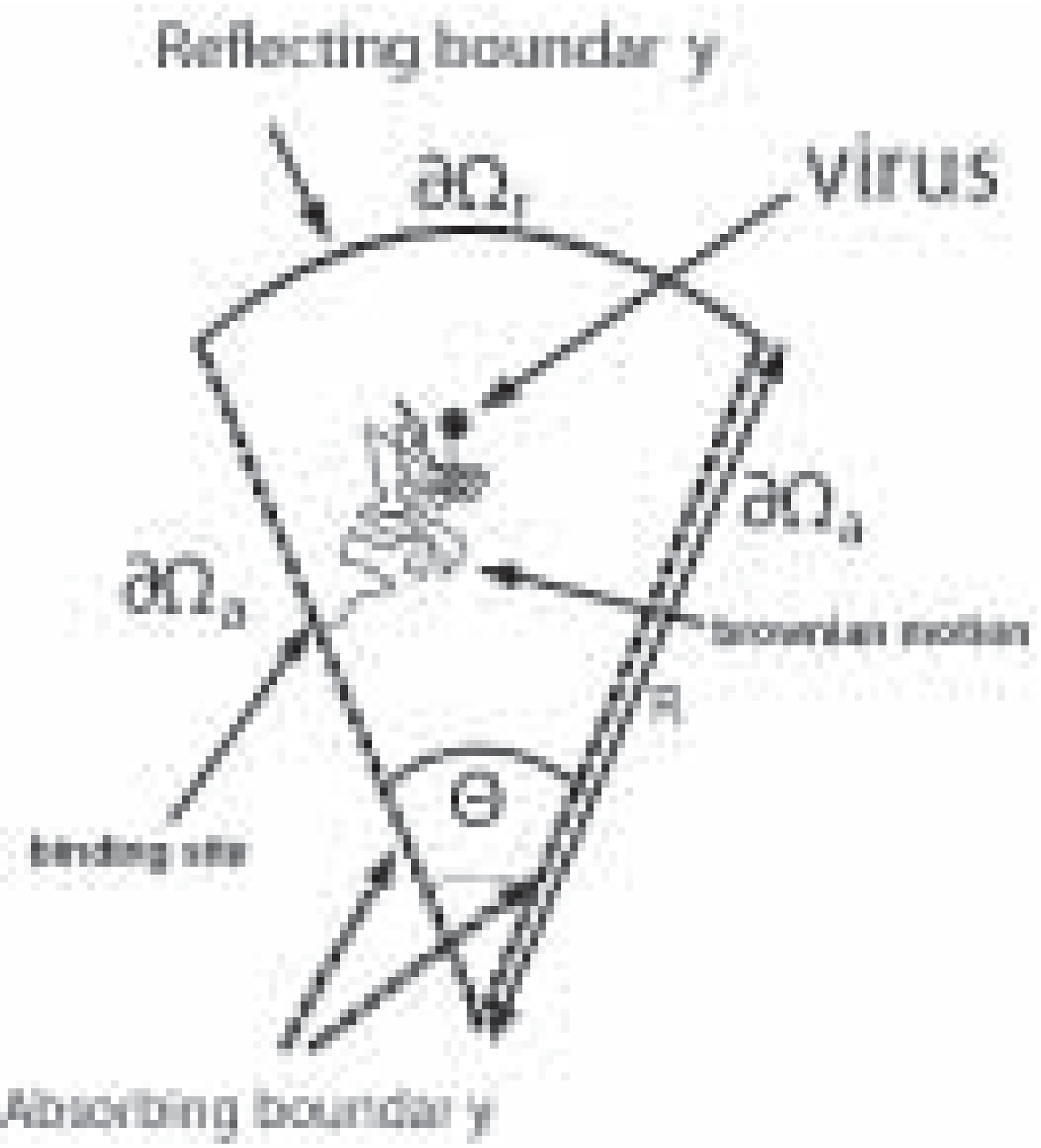,width=0.5\textwidth}}} \quad
 \mbox{ \subfigure[]{\epsfig{figure=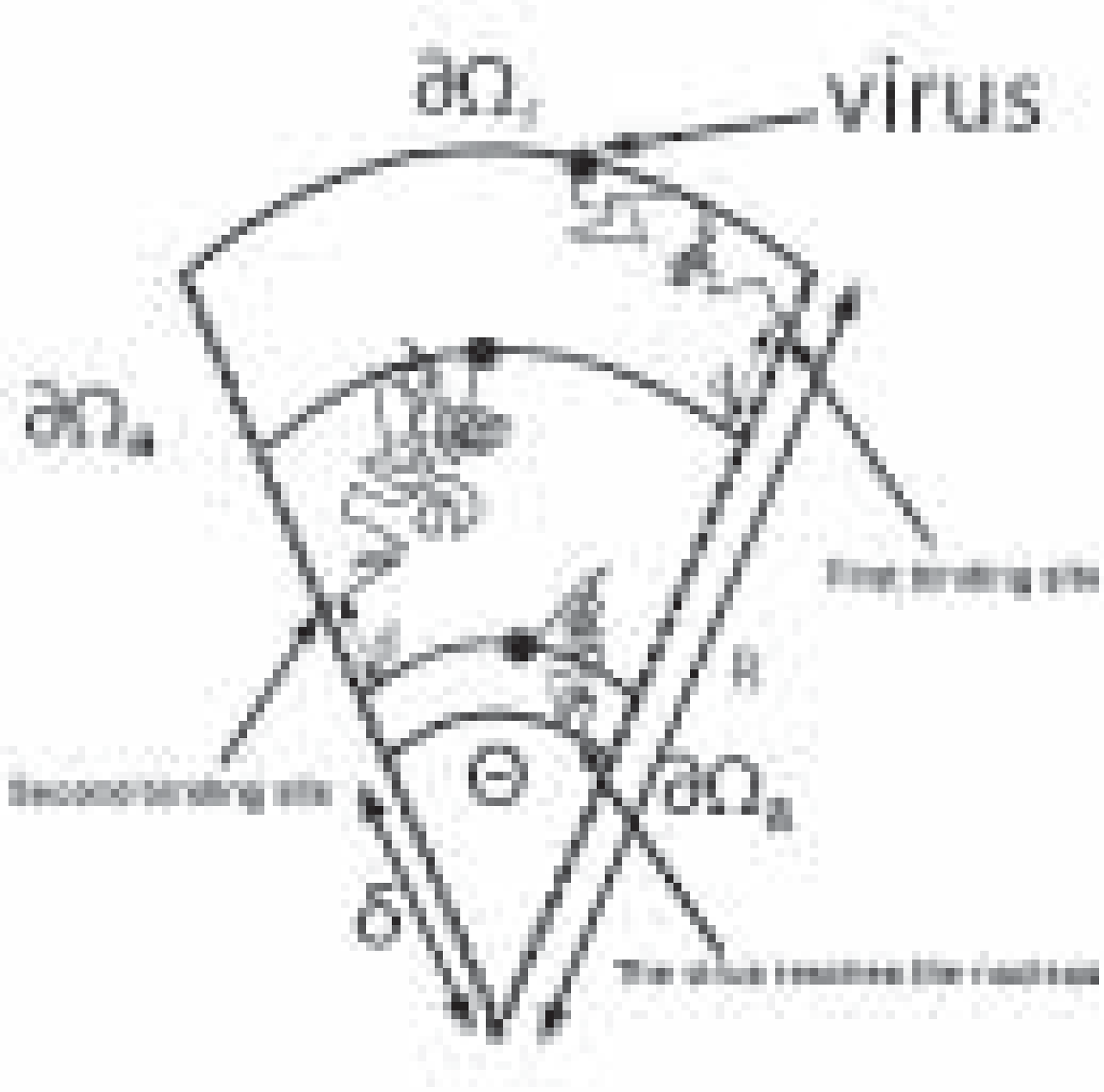,width=0.5\textwidth}}} \quad
\caption{{\bf Virus trafficking inside a cell.} (a) Representation of the cell portion between two
   microtubules. (b) Transport along microtubules: Two fundamental steps are represented. A fundamental step is made of the two
   intermediate step which are first the diffusion inside the domain
   followed by the directed motion along the microtubule.}
\label{iteration}
\end{figure}

Although a virus can drift along microtubules in both directions by
using dynein (resp. kinesin) motor proteins for the inward (resp.
forward) movement, we only take into account the drift toward the
nucleus \cite{Hirokawa}. It is still unclear what is the precise
mechanism used by a virus to select a direction of motion. Attached
to a dynein molecule, the virus transport consists in several steps
of few nanometers: the length of each step depends on the load of
the transported cargo and ATP-concentration \cite{Mallick}. We
neglect here the complexity of this process, assuming that ATP
molecules are abundant, uniformly distributed over the cell and is
not a limiting factor. We thus assume the bound particle moves towards the nucleus with the mean constant velocity $V$. When the particle is released away from the microtubule,
inside the domain, the process can start afresh  and the particle
diffuses freely. Because the Smoluchowski limit of the Langevin
equation does not account for the change in velocity, we release the
 the particle at a certain distance away from the
microtubule, but at a fixed distance from the nucleus (at an angle
chosen uniformly distributed), see figure
\ref{iteration}.

Because microtubules are taken uniformly distributed, we can always release
the virus inside the slice $\tilde{\Omega}$, between two neighboring
microtubules. Thus the movement of the virus will be studied in
$\tilde{\Omega}$: inside the cytosol, the viral movement is purely
Brownian until it hits a microtubule which is now the lateral
boundary of $\tilde{\Omega}$ (see figure (\ref{iteration})). We
assume that once a virus hits a microtubule, with probability one,
the dynamics switches from diffusion to a determinist motion with a
constant drift. A virus spends on a microtubule a time that we
consider to be exponentially distributed, since this time is the sum
of escape time from deep potential wells. We approximate the total
time on a microtubule by the mean time $t_m$. Thus a virus moves to a
distance $d_m=Vt_m$ along microtubule, which depends only on the
characteristic of the virus-microtubule interactions. To summarize,
the virus trajectory is a succession of diffusion steps mixed with
some periods of attaching and detaching to microtubules. Thus
scenario repeats until the virus hits the nucleus surface (Figure
(\ref{iteration})).

\subsection{Computing the MFPT to reach the nucleus}\label{descrip}
We define the {\em mean time to infection} as  the MFPT a virus
reaches the surface of the disk $D\left(\delta\right)$ inside the
domain $\tilde{\Omega}$ (see figure (\ref{iteration})).

To estimate the mean time to infection, we note that we can
decompose the overall motion as a repeated fundamental step. This
step consists of the free diffusion of the particle inside the
domain followed by the motion along the microtubule. The total time
of infection $\tau_i$ is then the sum of times the particle spends in
each step. Although the time on microtubule is determinist equal to
$t_m$, the diffusing time is not easy to compute and depend on the
initial condition. Ultimately $\tau_i$ depends on the number of
times the fundamental step repeats before the particle reaches the
nucleus.

Let us now described each step: the first step starts when the virus
enter the cell at the periphery $r=R=R_0$ (at a random angle $\theta
\in [0;\Theta]$) and ends when the virus hits either the lateral boundary or the nucleus.
We now consider the first passage time ${u}\left(R_0\right)$ to the
absorbing boundary and by $r(R_0)$ the hitting position. To account
for the determinist drift, we move during a deterministic time $t_m$
the virus from a distance $d_m$ along the microtubule. In that case,
the initial random position for the next step is given by $r =
R_1=r(R_0)-d_m$ and the total time in step $1$ is
${u}\left(R_0\right)+t_m.$

We iterate the process as follow and consider in each step k the
distance $R_k=r(R_{k-1})-d_m$ from which the particle starts and the
time  ${u}\left(R_k\right)+t_m$ it spends inside the step. If we
denote by $n_s$ the random number of steps necessary to reach the
nucleus $r=\delta$, the time to infection $\tau_i$  is given by
\beq \tau_i = \sum_{k=0}^{n_s-1} {u}(R_k) + n_s t_m + t_r \hbox{ ,}
\eeq
where $t_r$ is a residual time, which is the time to reach the
nucleus before a full step is completed.

We are interested in the estimating the mean first passage MFPT
$\tau$ of $\tau_i$, given by
\beq
\tau=E(\tau_i) = E\ds\left(\sum_{k=0}^{n_s-1} {u}(R_k)\right) + <n_s> t_m + <t_r>,
\eeq
where $<n_s>$ is the mean number of steps and $<t_r>$ is the mean
residual time. If we introduce the probability density function
$p_m=Pr\{n_s=m\}$ that the number of step is exactly equal to m, we
can write
\beq
\tau=E(\tau_i) = \sum_{m=1}^\infty E\ds\left(\sum_{k=0}^{n_s-1} {u}(R_k)|n_s=m\right)p_m + <n_s> t_m + <t_r>,
\eeq
To estimate the MFPT $\tau$, we shall approximate the previous sum
by using the mean first passage time $\bar{u}(R_k)$ in each step $k$.
To estimate $\bar{u}(R_k)$, we will solve (in the next paragraph)
the Dynkin's equation with the following boundary conditions: inside
$\tilde{\Omega}$, the particle is reflected at the periphery $r=R$,
absorbed at the nucleus $\p \tilde{\Omega}_a $ and at $\theta = 0$
and $\theta =\Theta$. We will also estimate the mean distance
$\bar{d}_k$ covered during step $k$. For that purpose we will estimate
the mean exit position $r_m(R_k)$, conditioned on the initial
position $r=R_k$. Indeed, we will thus get $\bar{d}_k=R_k-r_m(R_k)-d_m$. The estimates of the mean distances covered for each fundamental step will ultimately lead to an approximation of the mean number of step $n=<n_s>$: $n$ will be computed such that  $R_n \geq \delta$ and $R_{n+1} < \delta$ (where $R_n=r_m(R_{n-1})-d_m$ is defined recursively). Finally, we will obtain the following approximation for the infection time
\beq
\tau\approx \ds{\sum_{k=0}^{n-1} \bar{u}(R_k) + n t_m + <t_r>,}
\eeq
The mean residual time $<t_r>$ can  be equal either to $
\bar{u}(R_n)+\alpha t_m,$ where $0\leq \alpha<1$ if the virus binds to a
microtubule in the last step and travels a distance $\alpha d_m$ on
the microtubule, or to the MFPT to the nuclear boundary if
$r_m(R_n)<\delta$.

\section{Mean First Passage Time and Exit point distribution }
In first approximation, under the assumptions of a sufficiently
small radius $\delta<<R$ and an angle $\Theta<<1$ , for the
computation of the MFPT and the distribution of exit points, we
neglect the nuclear area. We define the full pie wedge $\Omega^R$
domain of angle $\Theta$. Inside $\Omega^R$, we use the boundary
conditions described above. Consequently, the  MFPT to a microtubule
$u = u\left(r,\theta\right)$ of a virus starting initially at
position $\left(r, \theta\right)$ is solution of the Dynkin's
equations
\cite{book}
\beq \label{edp1} \ds {
D \Delta u\left(\x\right) }&=& -1   \hbox{ for }  \x \in
\Omega^R
\\ & & \nonumber\\
u\left(\x\right)  &=& 0 \hbox{ for }  \x \in \p \Omega^R_a \nonumber \\
 & & \nonumber\\
\frac{\partial u}{\partial \n} &=& 0  \hbox{ for } \x \in \p
\Omega^R_r\hbox{ ,} \nonumber \eeq where $\p \Omega^R_a=\{\theta=0\}
\cup \{\theta=\Theta\}$ and $\Omega^R_r=
\{r=R\}$.
\subsection{The general solution for the MFPT}
In this paragraph only we reparametrize the domain by $-\Theta/2\leq
\theta\leq \Theta/2$. By writing equation (\ref{edp1}) in polar coordinates and using the
separation of variables, the general solution of equation
\beq \left(
\frac{\partial^2 u }{\partial r^2} +
\frac{1}{r}
\frac{\partial u}{\partial r} +
 \frac{1}{r^2} \frac{\partial^2 u}{\partial \theta^2}\right)\left(r,\theta\right) &=& -1  \hbox{ for } \left(r, \theta\right) \in \Omega^R   \\
u\left(r,\theta\right) &=& 0 \hbox{ for } \left(r,
\theta\right) \in \p\Omega^R_a \hbox{} .
\eeq
 is given by \cite{Redner} \beq \label{tau}
u\left(r,\theta\right) =
\frac{r^2}{4D}\left(\frac{cos\left(2\theta\right)}{cos\left(\Theta\right)}
- 1\right) + \sum_{n=0}^{\infty} A_n r^{\lambda_n}
cos\left(\lambda_n \theta\right), \hbox{ for } -\Theta/2\leq
\theta\leq \Theta/2
\eeq
where the edge boundary is {here} located at position $\theta=\pm
\Theta/2$. The sum in the right-hand
side is the general solution of the homogeneous problem $\Delta u =
0$ in $\Omega^R$. The boundary conditions on the sides of the wedge
impose that
\beq
\lambda_n =
\left(2n + 1\right)\frac{\pi}{\Theta}, \eeq while the reflecting
condition for $r=R$ reads \beq \label{bounR} \frac{\partial
u}{\partial r}\left(R, \theta\right) = 0 \hbox{ for all } \theta
\in [-\Theta/2,\Theta/2].
\eeq
Using the uniqueness of Fourrier decomposition and the boundary
condition (\ref{bounR}), we obtain that
\beq A_n  =
\frac{\left(-1\right)^{n+1}8 R^{2-\lambda_n}}{D\Theta
\lambda_n^2\left(\lambda_n^2 - 4\right)}.
\eeq
By averaging formula (\ref{tau}) over an initial uniform
distribution, the MFPT to a one of the wedge is given  by
\begin{equation} \label{tau1}
\bar{u}\left(r\right) = \frac{1}{\Theta}\int_{\theta =
0}^{\theta=\Theta} u\left(r,\theta\right) d\theta =
\frac{r^2}{4D}\left(\frac{tan \left(\Theta\right)}{\Theta} -
1\right) - \sum_{n=0}^{\infty}
\frac{16R^{2-\lambda_n}r^{\lambda_n}}{D\Theta^2\lambda_n^3\left(\lambda_n^2-4\right)},
\end{equation}
where $\lambda_n = \left(2n+1\right)\frac{\pi}{\Theta}$. For
$\Theta$ small, equation (\ref{tau1}) can be approximated by
\begin{equation}\label{time}
\ds{\bar{u}\left(r\right) = \frac{r^2}{4D}\left(\frac{ tan
\left(\Theta\right)}{\Theta} - 1\right) -\frac{16\Theta R^{2}
\left(\frac{r}{R}\right)^{\pi/\Theta}}{D\pi^3\left(\left(\pi/\Theta\right)^2-4\right)}}.
\end{equation}
\subsection{Exit points distribution}
To estimate the position a virus will attach preferentially to the
microtubule, we determine the distribution of exit points, when the
viral particle initially started at a radial distance from the
nucleus. We recall that the probability density function (pdf)
$p\left(\mathbf{r},t|\mathbf{r_0}\right)$ to find a diffusing
particle in a volume element $d\mathbf{r}$ at time t inside the
wedge $\tilde{\Omega}$, conditioned on the initial position
$\mathbf{r}=\mathbf{r_0}$ is solution of the diffusion equation
\beqq
\frac{\p p\left(\mathbf{r}, t|\mathbf{r_0}\right)}{\partial t} &=&
D\Delta p\left(\mathbf{r}, t| \mathbf{r_0}\right) \hbox{ for }
 \mathbf{r} \in \Omega^R  \\ & &\\
p\left(\mathbf{r}, t| \mathbf{r_0}\right) &=& 0  \hbox{ for } \mathbf{r} \in  \p \Omega^R_a \\ & &\\
\frac{\partial p\left(\mathbf{r}, t| \mathbf{r_0}\right)}{\partial
n} &=& 0 \hbox{ for } \mathbf{r} \in  \p \Omega^R_r \hbox{ ,}
\eeqq
where the initial condition is $p\left(\mathbf{r}, 0|
\mathbf{r_0}\right) = \delta\left(\mathbf{r} -
\mathbf{r_0}\right)$.
The distribution of exit points $\epsilon\left(\y\right)$ is given
by
\beq \epsilon\left(\y\right) = \ds{\int_0^{\infty}
j\left(\y,t\right)dt},
 \eeq
 where the flux $j$ is defined by
\begin{equation*} j\left(\y,t\right) = -D \frac{\partial
p\left(\mathbf{r},t\right)}{\partial \n }_{| \ds{\mathbf{r}} =\y}
\hbox{ .}
\end{equation*}
If we denote $C\left(\mathbf{r_0},\mathbf{r}\right) =
\int_0^{\infty} p\left(\mathbf{r}, t| \mathbf{r_0}\right) dt $
then C is solution of \beq \label{eqC} - D\Delta
C\left(\mathbf{r_0},\mathbf{r}\right)
=\delta\left(\mathbf{r}-\mathbf{r_0}\right) ,\eeq and \beq
\epsilon\left(\mathbf{y}\right) = -D\frac{\partial C}{\partial n}
\left(\mathbf{r_0},\mathbf{y}\right)   \hbox{ for }
 \mathbf{y} \in \Omega^R_a .
\label{eq:flux}
 \eeq
 Consequently, to obtain the pdf of exit points $\epsilon$,
we use the Green function in the wedge domain $\Omega^R$.  By
using a conformal transformation, we hereafter solve
a simplified case of an open wedge (\textit{i.e.} without a
reflecting boundary at $r=R$). This computation could be compared with the general one that will be derived in the next section.

To compute the exit points distribution, we consider the solution of
equation (\ref{eqC}), obtained by the image method and a conformal
transformation from the open wedge to the upper complex half-plane. The
Green function, solution of equation(\ref{eqC}) in the upper complex
half-plane is given by
 \beq C\left(z\right) = \frac{1}{2
\pi D} ln \frac{z-z_0}{z-z_0^{*}} ,
 \eeq
where $z_0^{*}$ the complex conjugate of $z_0$. Using the
conformal transformation $\omega =
f\left(z\right)=z^{\frac{\pi}{\Theta}}$ \cite{Henrici}, that maps
the interior of the wedge of opening angle $\Theta$ to the upper
half plane, the Green function in the wedge is given by \beq
C\left(z\right) = \frac{1}{2 \pi D}ln
\left(\frac{z^{\frac{\pi}{\Theta}} -
z^{\frac{\pi}{\Theta}}_0}{z^{\frac{\pi}{\Theta}} -
\left(z_0^{*}\right)^{\frac{\pi}{\Theta}}}\right).
\eeq
The flux to the line $\theta$ is given by \beq \epsilon_{\theta
}\left(r\right) = -\frac{D}{r} \frac {\partial C}{\partial \theta}
\left( re^{i\theta}\right) &=& \frac{1}{2\pi r} \frac {i\nu
\left(re^{i\theta}\right)^{\nu}.\left(k_0-k_0^{*}\right)}{\left(\left(re^{i\theta}\right)^{\nu}
- k_0\right)\left(\left(re^{i\theta}\right)^{\nu} -
k_0^{*}\right)} \nonumber \\
&=& \frac{1}{2\pi r} \frac {-2\nu \left(re^{i\theta}\right)^{\nu}
r_0^{\nu}
sin\left(\nu\theta_0\right)}{\left(re^{i\theta}\right)^{2\nu} +
r_0^{2\nu}-2\left(re^{i\theta}\right)^{\nu} r_0^{\nu} cos\left(\nu
\theta_0\right)}, \nonumber\eeq where $\nu = \frac{\pi}{\Theta}$,
$k_0 = z_0^{\nu} = \left(r_0 e^{i\theta_0}\right)^{\nu}$. Finally,
the exit point distribution for $\theta=\Theta$ is given by
\beq\label{eps1} \epsilon_{\Theta}\left(r\right) =
\frac{r_0}{\Theta} \frac { \left(r r_0\right)^{\left(\nu-1\right)}
sin\left(\nu\theta_0\right)}{r^{2\nu} + r_0^{2\nu} + 2\left(r
r_0\right)^{\nu} cos\left(\nu \theta_0\right)}, \eeq while for
$\theta=0$ it is given by
\begin{equation}
\epsilon_{0}\left(r\right) = \frac{r_0}{\Theta} \frac { \left(r
r_0\right)^{\left(\nu-1\right)}
sin\left(\nu\theta_0\right)}{r^{2\nu} + r_0^{2\nu} - 2\left(r
r_0\right)^{\nu} cos\left(\nu \theta_0\right)}.\label{distribopen}
\end{equation} A matlab check guarantees that
\begin{equation}
\int_0^{\infty}  \ds{\{\epsilon_{\Theta}\left(r\right) +
\epsilon_{0}\left(r\right)\} }dr = 1.
\end{equation}
This simple computation is instructive and shall be compared to the
full one given in section \ref{wedge}.

\subsection{Exit pdf in a Pie Wedge}\label{wedge}
To compute the exit points distribution in a pie wedge with a
reflecting boundary at $r=R$, we search for an explicit solution
of the diffusion equation in  polar coordinates inside the pie
wedge. We first consider the general diffusion equation
\begin{eqnarray} \label{diff}
\frac{\partial p}{\partial t}\left(\x,t|\y\right) &=&
D\left(\frac{\partial^2 p}{\partial r^2} +
\frac{1}{r}\frac{\partial p}{\partial r} +
\frac{1}{r^2} \frac{\partial^2 p}{\partial \theta^2}\right)\left(\x,t|\y\right)\\
p\left(\x,0| \y\right)&=& \delta\left(\x-\y\right) \nonumber
\end{eqnarray}
where the boundary conditions are given in (\ref{edp1}). We may
often use the change of variable $\forall n \in \mathbf{N}^{*}$ :
\begin{eqnarray*}
k = \frac{n\pi}{\Theta}.
\end{eqnarray*}
 The initial condition is given by
\begin{eqnarray*}
p\left(\x,0| \y\right)=p\left(r,\theta,0|r_0,\theta_0\right) =
\frac{2}{\Theta r_0} \delta\left(r-r_0\right) \sum_{k} sin
\left(k\theta\right) sin\left(k\theta_0\right),
\end{eqnarray*}
for $\theta<\theta_0$ (if $\theta>\theta_0$, $\theta_0$ must be
replaced  by $\Theta-\theta_0$). To compute the solution of equation
(\ref{diff}), we consider the Laplace transform $\hat{p}$ of the
probability p
\begin{eqnarray*}
s\hat{p}\left(r,\theta,s| r_0,\theta_0\right) - \frac{2}{\Theta
r_0} \delta\left(r-r_0\right) \sum_{k} sin \left(k \theta
\right)sin\left(k\theta_0\right) = D\left(\frac{\partial^2
\hat{p}}{\partial r^2} + \frac{1}{r}\frac{\partial
\hat{p}}{\partial r} + \frac{1}{r^2} \frac{\partial^2
\hat{p}}{\partial \theta^2}\right)\left(r,\theta,s|
r_0,\theta_0\right).
\end{eqnarray*}
Using the separation of variables, we have
\begin{eqnarray*}
\hat{p}\left(r,\theta,s| r_0,\theta_0\right)=\sum_{k}
R_k\left(r,s\right) sin \left(k\theta\right)
sin\left(k\theta_0\right),
\end{eqnarray*}
Using the change of variable, $x\left(s\right) =
r\sqrt{\frac{s}{D}}$ and $x_0\left(s\right) =
r_0\sqrt{\frac{s}{D}}$, we get for all k that
\begin{equation}
R_k^{''}\left(x\left(s\right),s\right) + \frac{1}{x\left(s\right)}
R_k^{'}\left(x\left(s\right),s\right) -
\left(1+\frac{k^2}{x\left(s\right)^2}\right)R_k\left(x\left(s\right),s\right)
= - \frac{2}{\Theta
Dx_0\left(s\right)}\delta\left(x\left(s\right)-x_0\left(s\right)\right).\label{bessel}
\end{equation}
$R_k\left(x\left(s\right),s\right)$ is a superposition of modified
Bessel functions of order $k$ :
$I_{k}\left(x\left(s\right)\right)$ and
$K_{k}\left(x\left(s\right)\right)$ for $x\left(s\right)\neq
x_0\left(s\right)$ :
\begin{eqnarray*}
R_k\left(x\left(s\right),s\right) = A_k
I_{k}\left(x\left(s\right)\right) + B_k
K_{k}\left(x\left(s\right)\right),
\end{eqnarray*}
where $A_k$ and $B_k$ are real constants. Since $K_{k}$ diverges
as $x\left(s\right) \rightarrow 0$, the interior solution for
$\left(x\left(s\right)<x_0\left(s\right)\right)$ depends only on
$I_{k}$. We denote by $D_{k}$ the exterior solution
for$\left(x\left(s\right)>x_0\left(s\right)\right)$. We use the
general notation $x\wedge y = min\left(x,y\right)$ and $x\vee y =
max\left(x,y\right)$, thus
\begin{eqnarray*}
R_k\left(x\left(s\right),s\right) =
A_kI_{k}\left(x\left(s\right)\wedge
x_0\left(s\right)\right)D_{k}\left(x\left(s\right)\vee
x_0\left(s\right)\right).
\end{eqnarray*}
To determine $D_{k} = a_k I_{k} + b_k K_{k}$, we use the
reflecting condition at $x\left(s\right) = x_{+}\left(s\right) =
R\sqrt{\frac{s}{D}}$ and we get that
\begin{eqnarray*}
A_k I_k\left(x_0\left(s\right)\right).\left(a_k
I^{'}_k\left(x_{+}\left(s\right)\right) + b_k
K^{'}_k\left(x_{+}\left(s\right)\right)\right) = 0 .
\end{eqnarray*}
We  choose
\begin{eqnarray*}
a_k = -K^{'}_k \left(x_{+}\left(s\right)\right) \hbox{ and } b_k =
I^{'}_k\left(x_{+}\left(s\right)\right).
\end{eqnarray*}
Thus
\begin{eqnarray*}
R_k\left(x\left(s\right),s\right) = A_k I_{k}\left(x\left(s\right)
\wedge x_0\left(s\right)
\right)\left(I^{'}_k\left(x_{+}\left(s\right)\right) K_k -
K^{'}_k\left(x_{+}\left(s\right)\right)I_k\right)\left(x\left(s\right)
\vee x_0\left(s\right)\right).
\end{eqnarray*}
The constants $A_k$ are determined by integrating equation
(\ref{bessel}) over an infinitesimal interval that includes $r_0$.
Using the continuity of $R_k$, we get
\begin{eqnarray*}
\left(R_k\right)^{'}_{x\left(s\right)>x_0\left(s\right)}|_{x\left(s\right)=x_0\left(s\right)}
-
\left(R_k\right)^{'}_{x\left(s\right)<x_0\left(s\right)}|_{x\left(s\right)=x_0\left(s\right)}
= - \frac{2}{\Theta Dx_0\left(s\right)},
\end{eqnarray*}
that is
\begin{eqnarray*}
A_k \left(I_k\left(I^{'}_k\left(x_{+}\left(s\right)\right)K_k^{'}
-K_k^{'}\left(x_{+}\left(s\right)\right)I^{'}_k\right)
 -
I^{'}_k\left(I^{'}_k\left(x_{+}\left(s\right)\right)K_k-K_k^{'}\left(x_{+}\left(s\right)\right)I_k\right)
\right)\left(x_0\left(s\right)\right) = - \frac{2}{\Theta
Dx_0\left(s\right)},
\end{eqnarray*}
after some simplifications, we get
\begin{eqnarray*}
A_k I^{'}_k\left(x_{+}\left(s\right)\right)\left(I_k
K_k^{'}-I^{'}_k K_k\right) \left(x_0\left(s\right)\right)= -
\frac{2}{\Theta Dx_0\left(s\right)}.
\end{eqnarray*}
Using the recurrent relation between modified Bessel functions (see
\cite{Abramowitz} or page 489 \cite{Carslaw}),
\begin{eqnarray*}
I^{'}_k\left(x_0\left(s\right)\right) =
\left(I_{k-1}-\frac{k}{x_0\left(s\right)}I_k\right)
\left(x_0\left(s\right)\right) \hbox{ and }
K^{'}_k\left(x_0\left(s\right)\right) =
\left(-K_{k-1}-\frac{k}{x_0\left(s\right)}K_k\right)\left(x_0\left(s\right)\right),
\end{eqnarray*}
we get
\begin{eqnarray*}
A_k I^{'}_k \left(x_{+}\left(s\right)\right) \left(
I_k\left(-K_{k-1}-\frac{k}{x_0\left(s\right)}K_k\right)-\left(I_{k-1}
-\frac{k}{x_0\left(s\right)}I_k\right)K_k\right)\left(x_0\left(s\right)\right)=
- \frac{2}{\Theta Dx_0\left(s\right)},
\end{eqnarray*}
that is
\begin{eqnarray*}
A_k I^{'}_k \left(x_{+}\left(s\right)\right) \left( I_k
K_{k-1}+I_{k-1}K_k \right) \left(x_0\left(s\right)\right)=
 \frac{2}{\Theta Dx_0\left(s\right)}.
\end{eqnarray*}
Finally, using this relation and the following Wronskian relation
(page 489
\cite{Carslaw}),
\begin{eqnarray*}
\left(I_{k}K_{k - 1} + I_{k - 1}K_{k}\right)
\left(x_0\left(s\right)\right) = \frac{1}{x_0\left(s\right)},
\end{eqnarray*}
we obtain that
\begin{eqnarray*}
A_k = \frac{2}{\Theta D I^{'}_k \left(x_{+}\left(s\right)\right)}.
\end{eqnarray*}
thus
\begin{eqnarray*} R_k\left(x\left(s\right),s\right) = \frac{2}{\Theta D
I^{'}_k\left(x_{+}\left(s\right)\right)}I_{k}\left(x\left(s\right)\wedge
x_0\left(s\right)\right)\left(I^{'}_k\left(x_{+}\left(s\right)\right)
K_k -
K^{'}_k\left(x_{+}\left(s\right)\right)I_k\right)\left(x\left(s\right)\vee
x_0\left(s\right)\right).
\end{eqnarray*}
We can now express the solution $\hat{p}$ for $\theta<\theta_0$ by
\begin{eqnarray*}
\hat{p}\left(r,\theta,s\right) = \frac{2}{\Theta D} \sum_k
\frac{I_{k}\left(x\left(s\right)\wedge
x_0\left(s\right)\right)\left(
I^{'}_k\left(x_{+}\left(s\right)\right) K_k -
K^{'}_k\left(x_{+}\left(s\right)\right)I_k
\right)\left(x\left(s\right)\vee
x_0\left(s\right)\right)}{I^{'}_k\left(x_{+}\left(s\right)\right)}sin\left(k\theta\right)sin\left(k\theta_0\right).
\end{eqnarray*}
The exit point distribution $\epsilon^0\left(r\right)$ is given by
\begin{eqnarray} \label{esp}
\epsilon^0\left(r\right)
=-\left(\frac{D}{r}\frac{\partial}{\partial
\theta}\left(\int_{0}^{\infty}p\left(r,\theta,t\right)dt\right)\right)\left(\theta=0\right).\label{exitdistrib}
\end{eqnarray}
To obtain an analytical expression for expression (\ref{esp}), we
use the Laplace relation:
\begin{eqnarray*}
\mathcal{L}\left(\int_{0}^{t} f\left(u\right) du\right) =
\frac{F\left(z\right)}{z},
\end{eqnarray*}
where $F = \mathcal{L}\left(f\right)$ is the Laplace transform of
the function $f$. We have
\begin{eqnarray*}
\int_0^t p\left(r,\theta,u\right) du
&=& \mathcal{L}^{-1}\left(\frac{\hat{p}\left(r,\theta,s\right)}{s}\right) \\
&=& \mathcal{L}^{-1}\left( \frac{2}{\Theta D}\sum_k
sin\left(k\theta\right)sin\left(k\theta_0\right)
\frac{I_{k}\left(x\left(s\right)\wedge
x_0\left(s\right)\right)\left(I^{'}_k\left(x_{+}\left(s\right)\right)
K_k -
K^{'}_k\left(x_{+}\left(s\right)\right)I_k\right)\left(x\left(s\right)\vee
x_0\left(s\right)\right)}{s
I^{'}_k\left(x_{+}\left(s\right)\right)}\right).
\end{eqnarray*}
The computation of the integral
\begin{equation}
I\left(r,\theta,t\right) = \frac{1}{\Theta \pi D i}\sum_k
sin\left(k\theta\right)sin\left(k\theta_0\right)\int_{-i\infty}^{+i\infty}
\frac{I_{k}\left(x\left(s\right)\wedge
x_0\left(s\right)\right)\left(I^{'}_k\left(x_{+}\left(s\right)\right)
K_k -
K^{'}_k\left(x_{+}\left(s\right)\right)I_k\right)\left(x\left(s\right)\vee
x_0\left(s\right)\right)}{s
I^{'}_k\left(x_{+}\left(s\right)\right)} e^{st}ds\label{integral}
\end{equation}
uses the residue theorem and the details are given in the Appendix.
We have
\begin{eqnarray*}
I\left(r,\theta,t\right) =\int_0^t p\left(r,\theta,u\right) du =
\frac{2}{\Theta D} \left(S_1(r,\theta,t) + S_2(r,\theta,t)\right),
\end{eqnarray*}
where
\begin{eqnarray*}
\ds{S_1(r,\theta,t)} &=& \ds{\sum_{k} sin\left(k\theta\right)sin\left(k\theta_0\right)
\frac{r^k\left(r_0^{2k} + R^{2k}\right)}{2kR^{2k}r_0^k},} \\
\ds{S _ 2(r,\theta,t)} &=& \ds{-2 \sum_{k}
sin\left(k\theta\right)sin\left(k\theta_0\right)
\sum_{j=1}^{\infty}e^{-D\alpha_{j,k}^2
t}\frac{J_k\left(r\alpha_{j,k}\right)J_k\left(r_0\alpha_{j,k}\right)}{\left(R^2\alpha_{j,k}^2
- k^2\right)J^2_k\left(R\alpha_{j,k}\right)}},
\end{eqnarray*}
and $J_k$ are the $k$-order Bessel's function and $\alpha_{j,k}$
are the roots of the equation:
\begin{eqnarray*}
J_k'\left(R \alpha\right) = 0.
\end{eqnarray*}
Consequently, for $r<r_0$, using (\ref{exitdistrib}), we get the
following exit distribution (for $\Theta = 0$) :
\begin{eqnarray*}
\epsilon^0\left(r\right) = \frac{2}{\Theta} \frac{\partial}{r
\partial \theta} \left(\lim_{t \rightarrow \infty} \left(S_1(r,\theta,t) +
S_2(r,\theta,t)\right)\right)_{\theta = 0}.
\end{eqnarray*}
Because :
\begin{equation*}
\lim_{t \rightarrow \infty} S_1(r,\theta,t) = S_1(r,\theta) \hbox{
and } \lim_{t \rightarrow \infty} S_2(r,\theta,t) = 0,
\end{equation*}
we finally obtain that
\begin{equation}
\epsilon^0\left(r\right) = \frac{1}{\Theta}\sum_{k}
sin\left(k\theta_0\right)\frac{r^{k-1}\left(r_0^{2k} +
R^{2k}\right)}{R^{2k}r_0^k},
\end{equation}
and, for $r>r_0$, a similar computation leads to :
\begin{equation}
\epsilon^0\left(r\right) = \frac{1}{\Theta }\sum_{k}
sin\left(k\theta_0\right)\frac{r_0^k\left(r^{2k} +
R^{2k}\right)}{R^{2k}r^{k+1}}.
\end{equation}
These expressions can be further simplified. Indeed, we rewrite them as follows (for $r<r_0$) :
\begin{eqnarray*}
\epsilon^0\left(r\right) = \frac{1}{\Theta r }\sum_{k}
sin\left(k\theta_0\right)\left(\frac{r}{r_0}\right)^k
\left(1+\left(\frac{r_0}{R}\right)^{2k}\right),
\end{eqnarray*}
{thus,
\begin{eqnarray*}
\epsilon^0\left(r\right) = \frac{1}{\Theta r } \Im  m\left(\sum_{n \geq 1 }
e^{i n\nu \theta_0}\left(\frac{r}{r_0}\right)^{n\nu}
\left(1+\left(\frac{r_0}{R}\right)^{2 n\nu}\right)\right),
\end{eqnarray*}
where $\Im m $ denotes the imaginary part of the expression. We
obtain two geometrical series that can be summed. We get:
\begin{equation*}
\epsilon^0\left(r\right) = \frac{1}{\Theta r } \Im m\left(\frac{e^{i\nu \theta_0} \left(\frac{r}{r_0}\right)^{\nu}}{1-e^{i\nu \theta_0} \left(\frac{r}{r_0}\right)^{\nu}}+\frac{e^{i\nu \theta_0} \left(\frac{r}{r_0}\right)^{\nu}\left(\frac{r_0}{R}\right)^{2\nu}}{1-e^{i\nu \theta_0} \left(\frac{r}{r_0}\right)^{\nu}\left(\frac{r_0}{R}\right)^{2\nu}}\right),
\end{equation*}
that is:
\begin{equation*}
\epsilon^0\left(r\right) = \frac{1}{\Theta r } \Im m\left(e^{i\nu \theta_0}\left(\frac{\left(\frac{r}{r_0}\right)^{\nu}}{1-e^{i\nu \theta_0} \left(\frac{r}{r_0}\right)^{\nu}} + \frac{\left(\frac{r r_0}{R^2}\right)^{\nu}}{1-e^{i\nu \theta_0} \left(\frac{r r_0}{R^2}\right)^{\nu}}\right)\right).
\end{equation*}
After some rearrangements, we obtain the following exit point
distribution on $\theta = 0$, conditioned on the initial position
$(r_0,\theta_0)$:
\begin{equation}
\epsilon^0(r) = \epsilon^0\left(r|r_0,\theta_0\right) =
\frac{1}{\Theta r} \left( \frac { \left(r
r_0\right)^{\nu}
sin\left(\nu\theta_0\right)}{r^{2\nu} + r_0^{2\nu} - 2\left(r
r_0\right)^{\nu} cos\left(\nu \theta_0\right)} + \frac { \left(r
r_0 R^2\right)^{\nu}
sin\left(\nu\theta_0\right)}{\left(rr_0\right)^{2\nu} + R^{4\nu} -
2\left(r r_0 R^2\right)^{\nu} cos\left(\nu \theta_0\right)}
\right), \label{00}
\end{equation}
for $0\leq r \leq R$. Similarly, for $\theta = \Theta$, we obtain
\begin{equation} \epsilon^{\Theta}\left(r\right)=\epsilon^{\Theta}\left(r|r_0,\theta_0\right) = \frac{1}{\Theta r}
\left( \frac { \left(r r_0\right)^{\nu}
sin\left(\nu\theta_0\right)}{r^{2\nu} + r_0^{2\nu} + 2\left(r
r_0\right)^{\nu} cos\left(\nu \theta_0\right)} + \frac { \left(r
r_0 R^2\right)^{\nu}
sin\left(\nu\theta_0\right)}{\left(rr_0\right)^{2\nu} + R^{4\nu} +
2\left(r r_0 R^2\right)^{\nu} cos\left(\nu \theta_0\right)}
\right). \label{theta}
\end{equation}
We notice that letting $R$ tends to $\infty$, we recover the expressions computed in the open wedge case ((\ref{eps1}) and (\ref{distribopen})).
\begin{figure}[ht]
   \begin{center}
      \includegraphics[width=1.0\textwidth]{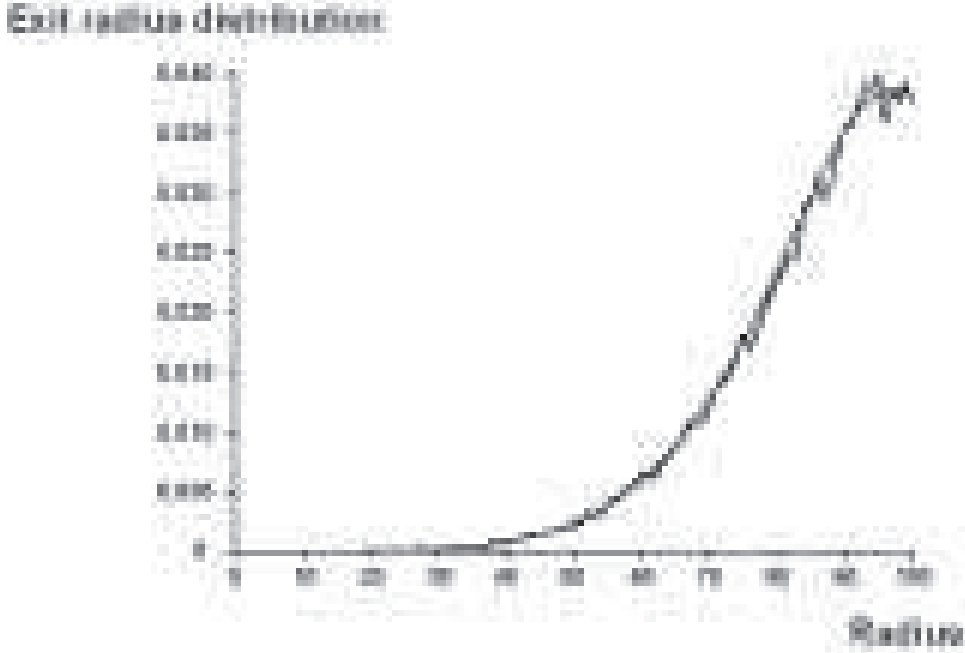}
   \end{center}
   \caption{{\bf Mean exit points distribution}. The theoretical distribution (dashed line) is tested against the empirical one (solid line) obtained by running a
simulation of 20 000 Brownian particles, starting on the wedge
bisectrix ($\theta_0 = \frac{\Theta}{2}$ at $r_0=R=100$ for
$\Theta=\frac{\pi}{6}$). Because the
starting point is located on the bisectrix,
$\epsilon^0\left(x\right) = \epsilon^{\Theta}\left(x\right),$ and
thus the analytical curve is given by $\epsilon\left(r\right) =
\epsilon^0\left(r\right)+\epsilon^{\Theta}\left(r\right)=\frac{2}{\Theta
r} \left( \frac { \left(r r_0\right)^{\left(\nu\right)}}{r^{2\nu}
+ r_0^{2\nu}} + \frac { \left(r r_0
R^2\right)^{\left(\nu\right)}}{\left(rr_0\right)^{2\nu} +
R^{4\nu}}\right)$. In that case, the maximum of the function
$\epsilon\left(r\right)$ is achieved at $r=r_0 e^{\frac{1}{2\nu}
ln\left(\frac{\nu-1}{\nu+1}\right)}.$}\label{exit points}
\end{figure}
\subsection{The Mean Exit Radius (MER)}
To determine the mean exit distribution radius
$\overline{\epsilon}\left(r|r_0\right)$ for a viral particle
starting initially at position $r_0,\theta_0$ where $\theta_0$ is
uniformly distributed between $0$ and $\Theta$, we consider
$\epsilon\left(r|r_0,\theta_0\right) =
\epsilon^0\left(r|r_0,\theta_0\right)
+\epsilon^\Theta\left(r|r_0,\theta_0\right)$ and estimate the
integral
\beq
\overline{\epsilon}\left(r|r_0\right) =
\frac{1}{\Theta}\int_{\Theta_0 =
0}^{\Theta}\epsilon\left(r|r_0,\theta_0\right) d\theta_0.
\eeq
 Integrating expressions ((\ref{00}) and (\ref{theta})) we get :
\begin{equation*}
\overline{\epsilon}\left(r|r_0\right) = \frac{2}{\Theta \pi r}
\left(ln\left(\frac{r^{\nu}+r_0^{\nu}}{|r^{\nu}-r_0^{\nu}|}\right)
+
ln\left(\frac{R^{2\nu}+\left(rr_0\right)^{\nu}}{R^{2\nu}-\left(rr_0\right)^{\nu}}\right)\right).
\end{equation*}
We define the  mean exit point as
$r_m\left(r_0\right)=\mathbf{E}\left(r|r_0\right)$ conditioned on
the initial radius $r_0$. Thus,
\beq
r_m\left(r_0\right)=\mathbf{E}\left(r|r_0\right) = \int_{0}^{R} r
\overline{\epsilon}\left(r|r_0\right) dr. \eeq
Using the expansion
$ln\left(1+x\right) =
\sum_{n\geq1}\left(-1\right)^{n+1}\frac{x^n}{n}$ for $x<1$, we obtain by a direct
integration that
\begin{equation}
r_m\left(r_0\right)=
\frac{8}{\pi^2}\left(r_0\left(\sum_{n=0}^{\infty}\frac{1}{\left(2n+1\right)^2}\left(\frac{1}{1-\frac{1}{\left(2n+1\right)^2\left(\frac{\pi}{\Theta}\right)^2}}\right)\right)
- R
\left(\sum_{n=0}^{\infty}\frac{\left(\frac{r_0}{R}\right)^{\left(2n+1\right)\frac{\pi}{\Theta}}
\frac{\pi}{\Theta}}{\left(2n+1\right)\left(\left(\left(2n+1\right)\frac{\pi}{\Theta}\right)^2-1\right)}\right)\right),
\end{equation}
using the expansion in the first part,
\beq
\frac{1}{1-\frac{1}{\left(2n+1\right)^2\left(\frac{\pi}{\Theta}\right)^2}}=
\sum_{p=0}^{\infty}\left(\frac{\Theta}{\left(2n+1\right)\pi}\right)^{2p}
\eeq
and the approximation $\Theta<<1$, we obtain using the value of the
Riemann $\zeta-$function, $\zeta\left(2\right)=\frac{\pi^2}{6}$ and
$\zeta\left(4\right)=\frac{\pi^4}{90}$, $r_0\leq R$, that
\begin{equation} \label{appr}
\ds{r_m\left(r_0\right) \approx
r_0\left(1+\frac{\Theta^2}{12}\right) -
\frac{8R}{\pi^2}\left(\frac{r_0}{R}\right)^{\pi/\Theta}
\frac{\pi/\Theta}{\left(\pi/\Theta\right)^2-1}}.
\end{equation}
For $\Theta$ small, the second term in the right-hand side of
(\ref{appr})  is exponentially small.
\section{Approximation of a virus motion by an effective Markovian stochastic equation}
We replace the successive steps of viral dynamics with an effective
stochastic equation containing a constant steady state drift.
\subsection{Methodology}
Virus motion described in paragraph (\ref{descrip}) consists of a
succession of drift and diffusing periods. We start with the
stochastic equation
\begin{eqnarray}\label{ou}
\mathbf{\dot{X}}= -B\frac{\mathbf{r}}{|\mathbf{r}|} +\sqrt{2D}\mathbf{\dot{w}},
\end{eqnarray}
where $\rr$ is the radial component of $\mathbf{X}$ , $B$ is the amplitude of the
drift. The MFPT of the process (\ref{ou}) to the nucleus located
$r=\delta$, when the initial position is located on the cell surface
$r=R$ is solution of
\begin{eqnarray*}
D\left(\frac{d^2t}{dr^2} +
\frac{1}{r}\frac{dt}{dr}\right)\left(r,\theta\right) -
B\frac{dt}{dr}
\left(r,\theta\right) &=& -1 \hbox { for } \left(r,\theta\right) \in \Omega   \\
t\left(r,\theta\right) &=&  0\hbox { for } r=\delta \\
\frac{d t}{dr}\left(r,\theta\right)&=&0 \hbox { for } r=R.
\end{eqnarray*}
A similar equation can be written in the domain $\tilde{\Omega}$
with reflective boundary conditions of the wedge. Both processes in
the full domain or in $\tilde{\Omega}$ lead to the same MFPT. The
solution $t(B,r)$ is given by
\begin{equation} \label{seol}
 t\left(B,r\right) = C-\int_{r}^R  \left(\int_{v}^R
 \frac{ue^{-\alpha\left(u-v\right)}}{Dv}du\right)dv,
\end{equation}
where $\alpha=\frac{B}{D}$ and
\begin{equation}
t\left(B,R\right)=C= \int_{\delta}^R
\left(\int_{v}^R
\frac{ue^{-\alpha\left(u-v\right)}}{Dv}du\right)dv.
\end{equation}
For a fixed radius R, the derivative of the function
$t\left(B,R\right)$ with respect to B is strictly negative, which
shows that $B\rightarrow t\left(B,R\right)$ is strictly decreasing.
To determine the value of the amplitude $B$, we equal the mean time
$t\left(B,R\right)$ with the MFPT to reach the nucleus within the
iterative procedure as described in paragraph (\ref{descrip}): at
time zero, the virus starts at a position $r=R=R_0$ and reaches the edge
boundary in a mean time $\bar{u}\left(R_0\right)$ and at a mean
position
 $r_m\left(R_0\right)$. The viral particle is then transported toward the nucleus over a distance $d_m$ during a
  time $t_m$. Either the particle reaches the nucleus before time $t_m$ and then the algorithm is terminated or
  in a second step, it starts at a position  $R_1=r_m\left(R_0\right)-d_m$. The
 process iterates until the particle reaches the nucleus. We consider the mean number of fundamental steps (diffusion step and directed motion along a MT step) the virus needs to reach the nucleus is equal to $n \geq 0$. The mean time to reach the nucleus computed by equation (\ref{seol}) has thus to be equal to the mean time $\tau=\sum_{k=0}^{n-1}
\bar{u}(R_k) + n t_m + <t_r>$ of the iterative trajectory.  In a first approximation, we neglect the mean residual time $<t_r>$ and we thus get the equality:
\begin{eqnarray}\label{final}
&&t\left(B,R\right) = \tau=
\sum_{k=0}^{n-1} \bar{u}\left(R_k\right) +
nt_m \\
&&R_{k+1} = r_m\left(R_k\right) - d_m \\
&&R_0=R\hbox{ .}
\end{eqnarray}
For a fix radius R, equation (\ref{final}) has a unique
solution B, which can be found in practice by any standard numerical
method.
\subsection*{Remark}
 The MFPT of a particle where the trajectory consists of alternating drift (traveling along microtubules)
 and diffusion periods can either be higher or lower than the MFPT of a pure Brownian particle. Indeed when
 $B<0$, the drift effect is less efficient than pure diffusion. For example,
 for $\Theta=\frac{\pi}{6}$, $R=100 \mu m$, $\delta=\frac{R}{4} = 25 \mu m$, a large diffusion
constant $D=10 \mu m^2 s^{-1}$ with the dynamical parameters $t_m=1s$ and $d_m=1 \mu m$, leads to a negative mean
drift
\beq
 B\approx-0.14 \mu m s^{-1}.
\eeq
 On the other hand, for a small diffusion constant $D=1 \mu m^2 s^{-1}$, an efficient microtubules
 transport obtained for $t_m=1s$ and $d_m=5 \mu m$ leads to a mean positive drift
\beq
 B\approx0.13 \mu m s^{-1}.
\eeq
\subsection{Explicit expression of the drift in the limit of $\Theta<<1$}
When the number of microtubules is large enough,  the condition
$\Theta<<1$ is satisfied. Moreover, because a virus entering a cell
surface has a deterministic motion, we can assume that the initial
position  satisfies $R_0<R$ so that we can neglect any boundary
effects and  use the open wedge approximation which consists of
using formula (\ref{appr}) without the boundary layer term.
Actually, this approximation is not that restrictive because after
the first iteration process (movement along the microtubule followed
by the particle release), the boundary layer term is negligible
compared to the other term.

To obtain an explicit expression for the amplitude B, we consider
the successive approximations
\begin{equation}
r_m\left(R_0\right)\approx R_0\left(1+\frac{\Theta^2}{12}\right),
\end{equation}
and
\begin{eqnarray*}
R_0 &=& R_0;\\
R_1 &\simeq& R_0\left(1+\frac{\Theta^2}{12}\right) - d_m;\\
R_2 &\simeq& R_0\left(1+\frac{\Theta^2}{12}\right)^2 - d_m\left(1+\left(1+\frac{\Theta^2}{12}\right)\right);\\
\vdots\\
R_i &\simeq& R_0\left(1+\frac{\Theta^2}{12}\right)^i - d_m\left(\sum_{k=0}^{i-1}\left(1+\frac{\Theta^2}{12}\right)^k\right);\\
\end{eqnarray*}
that is
\begin{equation}
R_i \simeq \left(R_0-\frac{12
d_m}{\Theta^2}\right)\left(1+\frac{\Theta^2}{12}\right)^{i} +
\frac{12 d_m}{\Theta^2}.
\end{equation}
Thus  the particle reaches the nucleus after $n$ iteration steps
which {approximatively} satisfies  $R_n=\delta$,
\begin{equation}
n \simeq \ds{\frac{ln\left(\frac{1-\frac{\delta\Theta^2}{12
d_m}}{1-\frac{R_0\Theta^2}{12
d_m}}\right)}{ln\left(1+\frac{\Theta^2}{12}\right)}}\approx
\frac{R_0-\delta}{d_m} +o\left(1\right).
\end{equation}
If $T_n$ denotes the mean time a viral particle takes to reach the
nucleus, then using formula (\ref{time}), we obtain
\begin{equation}
T_n \simeq n.t_m +
\frac{\left(\frac{tan\left(\Theta\right)}{\Theta}-1\right)}{4D}\sum_{i=0}^{n-1}R_i^2
\hbox{ ,}
\end{equation}
that is
\begin{eqnarray*}
t &\simeq& n.t_m +
\frac{\left(\frac{tan\left(\Theta\right)}{\Theta}-1\right)}{4D}\\
&&\sum_{i=0}^{n-1} \left(\left(\frac{12 d_m}{\Theta^2}\right)^2 +
2\left(\frac{12 d_m}{\Theta^2}\right) \left(R_0-\frac{12
d_m}{\Theta^2}\right)
\left(1+\frac{\Theta^2}{12}\right)^{i}+\left(R_0-\frac{12
d_m}{\Theta^2}\right)^2\left(1+\frac{\Theta^2}{12}\right)^{2i}\right),
\end{eqnarray*}
\begin{eqnarray*}
T_n &\simeq& n t_m +
\frac{\left(\frac{tan\left(\Theta\right)}{\Theta}-1\right)}{4D}\\
&&\left( n\left(\frac{12 d_m}{\Theta^2}\right)^2 - \left(\frac{24
d_m}{\Theta^2}\right) \left(R_0-\frac{12 d_m}{\Theta^2}\right)
\frac{1-\left(1+\frac{\Theta^2}{12}\right)^n}{\frac{\Theta^2}{12}}
+ \left(R_0-\frac{12 d_m}{\Theta^2}\right)^2
\frac{1-\left(1+\frac{\Theta^2}{12}\right)^{2n}}{1-\left(1+\frac{\Theta^2}{12}\right)^2}
\right).
\end{eqnarray*}
For $\Theta<<1$, a Taylor expansion gives that
\begin{eqnarray*}
T_n &\simeq& \left(\frac{R_0-\delta}{d_m}\right) t_m + \frac{
t_m\left(R_0-\delta\right)}{24 d_m}\left(1+\frac{R_0+\delta}{d_m}\right)\Theta^2\\
&+& \frac{\left(R_0-\delta\right)}{72
D}\left(d_m+3\left(R_0+\delta\right)+
\frac{2\left(R_0^2+R_0\delta+\delta^2\right)}{d_m}\right)\Theta^4
+ o\left(\Theta^4\right).
\end{eqnarray*}
In small diffusion limit  $D<<1,\Theta<<1$, the velocity is  $B
\simeq
\frac{R_0-\delta}{T_n}$ and consequently we obtain for $R_0 \approx R$, a second order approximation
\begin{equation}
B\approx\frac{\frac{d_m}{t_m}}{1+\left(1+\frac{R+\delta}{d_m}\right)\frac{\Theta^2}{24}
+ O\left(\Theta^4\right)},
\end{equation}
where $d_m, {t_m}$ are the mean distance and the mean time a virus
stays on the microtubule, R (resp. $\delta$) is the radius of the
cell (resp. nucleus) and  $\Theta=\frac{2\pi}{N}$, where N is the
total number of microtubules.

\subsection{Justification of the MFPT-criteria.}

To justify the use of the  MFPT-criteria to estimate the steady
state drift, we run numerical simulations of 1,000 viruses inside a
 two dimensional domain $\Omega$ ($\delta<r<R$) with intermittent dynamics,
 alternating between epochs of free diffusion and directed motion along microtubules and compare the
steady state distribution with the one obtained by solving the
Fokker-Planck equation for viruses whose trajectories are described
by the effective stochastic equation
(\ref{eq2}) with our computed constant drift
\beq
\mathbf{b}
\left(\mathbf{X}\right)=-\frac{\frac{d_m}{t_m}}{1+\left(1+\frac{R+\delta}{d_m}\right)\frac{\Theta^2}{24}}
\frac{\mathbf{r}}{|\mathbf{r}|}=-B\frac{\mathbf{r}}{|\mathbf{r}|}.
\eeq
We imposed reflecting boundary conditions at the nuclear and the
external membrane. The theoretical normalized steady state
distribution $\rho$ satisfies
\begin{eqnarray*}
 D\Delta \rho-\nabla .[\mathbf{b} \rho] &=& 0 \hbox{ in } \Omega\\
 \frac{d \rho}{dr}\left(R\right)=\frac{d \rho }{dr}\left(\delta\right)&=&0.
 \end{eqnarray*}
and the solution $\rho$ is given by
 \beq
 \rho(r)=\frac{e^{-\frac{Br}{D}}}{\int_{\delta}^{R}e^{-\frac{Br}{D}} 2\pi r dr}=\frac{e^{-\frac{Br}{D}}}{2\pi \frac{D}{B}\left(\delta e^{-\frac{B\delta}{D}} - R e^{-\frac{B R}{D}}+\frac{D}{B}\left(e^{-\frac{B \delta}{D}}-e^{-\frac{B R}{D}}\right)\right)}.
 \eeq
The result of both distributions is presented in figure \ref{compa}
where we can observe that both curves match very nicely.
This result shows that the criteria we have used is at least enough
to recover the distribution. For the simulations, we consider the directed run of the virus along a MT (loaded by dynein) lasts $t_m=1s$ and covers a mean distance $d_m=0.7\mu m$  \cite{schroer}. The diffusion
constant is $D=1.3\mu m^2 s^{-1}$ as observed for the Adeno
Associated Virus
\cite{Seisenberger}.
 \begin{figure}[ht]
   \begin{center}
      \includegraphics[width=1.0\textwidth]{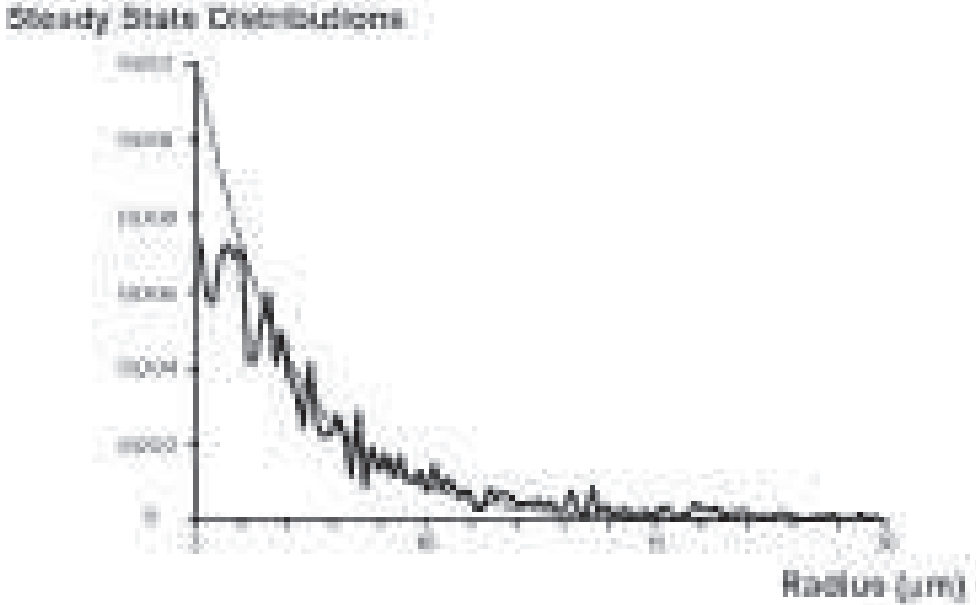}
   \end{center}
   \caption{{\bf
Steady State distributions.} We show the empirical steady state
distribution for $1,000$ viral trajectories with an intermittent
dynamic (solid line). The theoretical distribution of viruses whose
trajectories are described by the stochastic equation (\ref{eq2}) is
given in dashed line. Geometrical parameters are : $R=20
\mu m$, $\delta=5\mu m$ and $\Theta=\frac{\pi}{24}$.}\label{compa}
\end{figure}
The two curves in figure \ref{compa} fit very nicely except at the
neighborhood of the nuclear membrane, where the simulation of the
empirical distribution is plagued with a possible boundary layer.
Another source of discrepancy comes from the difference of behavior
of viruses far and close to the nucleus: viruses far from the nucleus do
not bind as often as those located in its neighborhood.
Consequently, a constant effective drift cannot account for the
radial geometry near the nucleus. A theory for radius dependent
effective drift has been derived in
\cite{LagacheHolcman}.

\section{Conclusion}
In the limit of a cell containing an excess of microtubules, we have
presented here a model to describe the motion of biological
particles such as viruses, vesicles and many others moving inside
the cell cytoplasm by a complex combination of Brownian motion and
deterministic drift. Our procedure consists mainly in approximating
an alternative switching mode between diffusion and deterministic
drift epochs by a steady state stochastic equation. This procedure
consists of estimating the amplitude of the effective drift and is
based on the criteria that the MFPTs to the nucleus, computed in
both cases are equaled. In that case, this amplitude account for the
directed transport along microtubules, the cell geometry and the
binding constants.
The model has however several limitations. First, we do not take
into account directly the backward movement of the virus along the
microtubules \cite{Sodeik1,Gross}, which can affect the mean time
and the amplitude of the drift. Second, the present computations are
given for two dimensional cell geometry only. It can still be
applied to many {\it in vitro} culture cells, however it is not
clear how to generalize our approach to a three dimensional cell
geometry.  For example, to study the trafficking inside cylindrical
axons or dendrites of neuronal cells, a different approach should
include this geometrical features. However despite these real
difficulties, the present model may be used to analyze plasmid
transport in an host cell, at the molecular level, which is one of
the fundamental limitation of gene delivery
\cite{Whittaker,Dean,Campbell,Luo}.

\section*{Appendix}
In this appendix, we provide an explicit computation of integral
(\ref{integral}) using the method of the residues. This method was
previously used in a similar context in (\cite{Carslaw} p 386). We
denote by $\left(p_j^k\right)_{j\geq 0}$ the poles of the function
\begin{eqnarray*}
\Phi:s\rightarrow \frac{I_{k}\left(x\left(s\right)\wedge
x_0\left(s\right)\right)\left(I^{'}_k\left(x_{+}\left(s\right)\right)
K_k -
K^{'}_k\left(x_{+}\left(s\right)\right)I_k\right)\left(x\left(s\right)\vee
x_0\left(s\right)\right)}{s
I^{'}_k\left(x_{+}\left(s\right)\right)} e^{st}.
\end{eqnarray*}
where ($x\left(s\right)=r\sqrt{\frac{s}{D}}$ , $x_0\left(s\right)
= r_0\sqrt{\frac{s}{D}}$ and $x_{+}\left(s\right) =
R\sqrt{\frac{s}{D}})$. The associated residues are
$\left(r_j^k\right)_{j\geq 0}$. We now compute the residues
explicitly.

To identify the poles, we recall the relation between the $k$-order
Bessel's function  $J_k$  (that is true for $z$ such that
$-\pi<arg\left(z\right)<\frac{\pi}{2}$) and the modified Bessel
functions $I_k$ (p 375
\cite{Abramowitz}):
\begin{equation}
I_k\left(z\right) = e^{-\frac{1}{2}k \pi
i}J_{k}\left(ze^{\frac{1}{2}\pi i}\right) \label{identity}.
\end{equation}
All roots $\alpha_{j,k}$ of the equations
\begin{eqnarray*}
J^{'}_k\left(R \alpha\right) = 0,
\end{eqnarray*}
 are real, simple and strictly positive (p 370
\cite{Abramowitz}) because  $k$ is real and
\begin{equation*}
k\leq\alpha_{1,k}<\alpha_{2,k} \ldots
\end{equation*}
Thus,
\begin{eqnarray*}
I^{'}_k\left(-iR\alpha_{j,k}\right) = 0.
\end{eqnarray*}
Finally the poles of $\Phi$ are simple given by $p_0^k=0$ and
$\forall j\geq 1$, $p_j^k = -D\alpha_{j,k}^2$. Consequently the
associated residues are given for each $k$  for all $j\geq 0$ by
\begin{equation}
 r_j^k = \lim_{s \rightarrow p_j^k}
\left(s-p_j^k\right) \Phi(s).\label{limit}
\end{equation}
Then using the residues, integral (\ref{integral}) is given by
\begin{eqnarray*}
I\left(r,\theta,t\right) = \frac{1}{\Theta \pi D i}\sum_k
sin\left(k\theta\right)sin\left(k\theta_0\right)\left(2\pi
i\right)\sum_{j \geq 0} r_j^k = \frac{2}{\Theta D}\sum_k
sin\left(k\theta\right)sin\left(k\theta_0\right)\sum_{j \geq 0}
r_j^k .\label{residtheorem}
\end{eqnarray*}
We now compute the residues $r_j^k.$ The residue $r_0^k$ is
associated with the pole $p_0^k=0$ and given by
\begin{eqnarray*}
r_0^k = \lim_{s\rightarrow 0} s \Phi(s)
\end{eqnarray*}
Using the following identities  on the modified Bessel functions (p
489 \cite{Carslaw})
\begin{eqnarray*}
I^{'}_k\left(z\right) =
I_{k+1}\left(z\right)+\frac{k}{z}I_k\left(z\right) \hbox{ and }
K^{'}_k\left(z\right) =
-K_{k-1}\left(z\right)-\frac{k}{z}K_k\left(z\right),
\end{eqnarray*}
substituting the derivatives $I^{'}_k$ and $K^{'}_k$ in the
expression of $\Phi$, we get
\begin{eqnarray*}
r_0^k &=&  \lim_{s\rightarrow 0}
\frac{I_{k}\left(x\left(s\right)\wedge
x_0\left(s\right)\right)}{\left(I_{k+1}+
\frac{k}{x_{+}\left(s\right)}I_k\right)\left(x_{+}\left(s\right)\right)}
\\
&&\left( \left( \left(I_{k+1}+\frac{k}{x_{+}\left(s\right)}I_k
\right)\left(x_{+}\left(s\right)\right)K_k\right)
 +
 \left(
\left(K_{k-1}+\frac{k}{x_{+}\left(s\right)}K_k
\right)\left(x_{+}\left(s\right)\right)I_k\right)\right)
\left(x\left(s\right)\vee x_0\left(s\right)\right),
\end{eqnarray*}
Taking into account the dominant terms only, we get
\begin{eqnarray*}
r_0^k = \lim_{s\rightarrow 0}
\frac{I_{k}\left(x\left(s\right)\wedge
x_0\left(s\right)\right)\left(
I_k\left(x_{+}\left(s\right)\right)K_k +
K_k\left(x_{+}\left(s\right)\right)
I_k\right)\left(x\left(s\right)\vee
x_0\left(s\right)\right)}{I_k\left(x_{+}\left(s\right)\right)}.
\end{eqnarray*}
To further compute this limit, we use the Taylor expansions of $I_k$
and $K_k$ (p 375 \cite{Abramowitz}) expressed in terms of the
$\Gamma$ function:
\begin{eqnarray*}
I_k\left(z\right) &\approx& \frac{\left(\frac{1}{2}
z\right)^{k}}{\Gamma\left(k+1\right)}\hbox{ and } K_k\left(z\right)
\approx
\frac{1}{2}\Gamma\left(k\right)\left(\frac{1}{2} z\right)^{-k}.
\end{eqnarray*}
For $r<r_0$, we get
\begin{eqnarray*}
r_0^k = \lim_{s\rightarrow 0} \frac{\frac{\left(\frac{1}{2}
\left(x\left(s\right)\right)\right)^{k}}{\Gamma\left(k+1\right)}
\left( \frac{\left(\frac{1}{2}
\left(x_{+}\left(s\right)\right)\right)^{k}}{\Gamma\left(k+1\right)}
\frac{1}{2}\Gamma\left(k\right)\left(\frac{1}{2}
\left(x_0\left(s\right)\right)\right)^{-k} +
\frac{1}{2}\Gamma\left(k\right)\left(\frac{1}{2}
\left(x_{+}\left(s\right)\right)\right)^{-k}
\frac{\left(\frac{1}{2}
\left(x_0\left(s\right)\right)\right)^{k}}{\Gamma\left(k+1\right)}\right)}{\frac{\left(\frac{1}{2}
\left(x_{+}\left(s\right)\right)\right)^{k}}{\Gamma\left(k+1\right)}}.
\end{eqnarray*}
Finally, using the relation $\Gamma\left(k+1\right) = k
\Gamma\left(k\right)$, and the expressions of $x(s)$, $x_0(s)$ and
$x_{+}(s)$ we get
\begin{eqnarray*}
r_0^k = \frac{r^k\left(r_0^{2k} + R^{2k}\right)}{2kR^{2k}r_0^k}.
\end{eqnarray*}
The computation of the other residues $\left(r_j^k\right)_{j \geq
1}$, is slightly different
\begin{eqnarray*}
r_j^k = \lim_{s\rightarrow p_j^k} \left(s-p^k_j\right)\Phi(s),
\end{eqnarray*}
where $p_j^k = -D\alpha^2_{j,k}$. Using the Wronskian relation (p
489 \cite{Carslaw}) :
\begin{equation*}
I_k\left(z\right)K'_k\left(z\right)-K_k\left(z\right)I'_k\left(z\right)=-\frac{1}{z},
\end{equation*}
we now substitute
\begin{eqnarray*}
K_k^{'}\left(z\right)=\frac{-\frac{1}{z} +
K_k\left(z\right)I^{'}_k\left(z\right)}{I_k\left(z\right)}.
\end{eqnarray*}
in the expression of $\Phi$, we get
\begin{eqnarray*}
r_j^k = \lim_{s\rightarrow p_j^k} \frac{\left(s-p_j^k\right)
e^{st}}{s}
\frac{I_{k}\left(x\left(s\right)\right)\left(I^{'}_k\left(x_{+}\left(s\right)\right)
K_k - \left(\frac{-\frac{1}{x_{+}\left(s\right)} + K_k
I^{'}_k}{I_k}\right)\left(x_{+}\left(s\right)\right)I_k\right)\left(x_0\left(s\right)\right)}{
I^{'}_k\left(x_{+}\left(s\right)\right)}.
\end{eqnarray*}
Because
\begin{equation*}
\lim_{s\rightarrow p_j^k} I^{'}_k\left(x_{+}\left(s\right)\right)
= I^{'}_k\left(x_{+}\left(p_j^k\right)\right) = 0,
\end{equation*}
we obtain the expression for the residues:
\begin{eqnarray*}
r_j^k = \frac{e^{p_j^k t}}{p_j^k}
\frac{I_k\left(x\left(p_j^k\right)\right)I_k\left(x_0\left(p_j^k\right)\right)}{I_k\left(x_{+}\left(p_j^k\right)\right)x_{+}\left(p_j^k\right)}
\lim_{s\rightarrow p_j^k}
\frac{\left(s-p^k_j\right)}{I^{'}_k\left(x_{+}\left(s\right)\right)}.
\end{eqnarray*}
Finally, since
\begin{equation*}
\lim_{s\rightarrow p_j^k}
\frac{\left(s-p^k_j\right)}{I^{'}_k\left(x_{+}\left(s\right)\right)}
= \frac{2\sqrt{D p_j^k}}{R} \lim_{s\rightarrow p_j^k}
\frac{x_{+}\left(s\right)-x_{+}\left(p_j^k\right)}{I^{'}_k\left(x_{+}\left(s\right)\right)
- I^{'}_k\left(x_{+}\left(p_j^k\right)\right)} = \frac{2\sqrt{D
p_j^k}}{R I^{''}_k\left(x_{+}\left(p_j^k\right)\right)},
\end{equation*}
we obtain
\begin{eqnarray*}
r_j^k = \frac{e^{p_j^k t}}{p_j^k}
\frac{I_k\left(x\left(p_j^k\right)\right)I_k\left(x_0\left(p_j^k\right)\right)}{I_k\left(x_{+}\left(p_j^k\right)\right)
x_{+}\left(p_j^k\right)} \frac{2\sqrt{Dp_j^k}}{R
I^{''}_k\left(x_{+}\left(p_j^k\right)\right)}.
\end{eqnarray*}
To simplify this expression,  we use that $I_k$ satisfies the
differential equation (p 374 \cite{Abramowitz}):
\begin{eqnarray*}
I^{''}_k\left(z\right) + \frac{1}{z} I^{'}_k\left(z\right) -
\left(1+\frac{k^2}{z^2}\right)I_k\left(z\right) = 0,
\end{eqnarray*}
thus for $z = x_{+}\left(p_j^k\right)$ :
\begin{eqnarray*}
I^{''}_k\left(x_{+}\left(p_j^k\right)\right) = \frac{p_j^k R^2 + D
k^2}{p_j^k R^2} I_k\left(x_{+}\left(p_j^k\right)\right),
\end{eqnarray*}
we get
\begin{equation*}
r_j^k = \frac{2De^{p_j^k t}}{R^2 p_j^k + D k^2}
\frac{I_k\left(x\left(p_j^k\right)\right)I_k\left(x_0\left(p_j^k\right)\right)}{I^2_k\left(x_{+}\left(p_j^k\right)\right)},
\end{equation*}
and finally, using (\ref{identity}), we get
\begin{eqnarray*}
r_j^k = \frac{2 e^{-D \alpha^2_{j,k} t}}{- R^2 \alpha^2_{j,k} +
k^2}
\frac{J_k\left(r\alpha_{j,k}\right)J_k\left(r_0\alpha_{j,k}\right)}{J^2_k\left(R\alpha_{j,k}\right)}.
\end{eqnarray*}
Integral (\ref{integral}) is given by
\begin{equation}
I(r,\theta,t) = \frac{2}{\Theta D}\sum_k
sin\left(k\theta\right)sin\left(k\theta_0\right)\sum_{j \geq 0}
r_j^k = \frac{2}{\Theta D} \left(S_1(r,\theta,t) +
S_2(r,\theta,t)\right).
\end{equation}
where
\begin{eqnarray*}
\ds{S_1(r,\theta,t)} &=& \ds{\sum_{k}
sin\left(k\theta\right)sin\left(k\theta_0\right)
\frac{r^k\left(r_0^{2k} + R^{2k}\right)}{2kR^{2k}r_0^k},} \\
\ds{S _ 2(r,\theta,t)} &=& \ds{-2 \sum_{k}
sin\left(k\theta\right)sin\left(k\theta_0\right)
\sum_{j=1}^{\infty}e^{-D\alpha_{j,k}^2
t}\frac{J_k\left(r\alpha_{j,k}\right)J_k\left(r_0\alpha_{j,k}\right)}{\left(R^2\alpha_{j,k}^2
- k^2\right)J^2_k\left(R\alpha_{j,k}\right)}},
\end{eqnarray*}

\section*{Acknowledgments}
 D. H. is partially supported by the program ``Chaire d'Excellence''
 from the French Ministry of Research.


\begin{thebibliography}{14}

\bibitem{Wiethoff} {\sc Wiethoff C.~M. Wiethoff and C.~R. Middaugh}, {\em Barriers to Non-Viral Gene
Delivery}, Journal of Pharmaceutical  Sciences, 92 (2003), pp.~203--217.

\bibitem{Dauty} {\sc D. Dauty and A.~S. Verkman}, {\em Actin Cytoskeleton as the Principal
Determinant of Size-Dependent DNA Mobility in Cytoplasm: a New
Barrier for Non-Viral Gene Delivery}, Journal of Biological Chemistry, 280 (2005), pp.~7823--7828. 


\bibitem{Dinh} {\sc A.~T. Dinh, T. Theofanous and S. Mitragotri}, {\em A Model for
Intracellular Trafficking of Adenoviral Vectors}, Biophysical Journal, 89 (2005), pp.~1574--1588.


\bibitem{alberts} {\sc B. Alberts, A. Johnson, J. Lewis, M. Raff , K. Roberts and P.
Walter}, {\em Molecular Biology of the Cell}, 4th Edition, Garland, New-York 2002.

\bibitem{david} {\sc D. Holcman}, {\em Modeling Trafficking of a Virus and a DNA Particle in the Cell Cytoplasm}, Journal of Statistical Physics, 127 (2007), pp.~471--494.

\bibitem{book} {\sc Z. Schuss}, {\em Theory and Applications of Stochastic
 Differential Equations}, John Wiley \& Sons Inc, New-York  1981.

\bibitem{Hirokawa} {\sc N. Hirokawa},
{\em Kinesin and Dynein Superfamily Proteins and the Mechanism of
Organelle Transport}, Science, 279 (1998), pp.~519--526.
\bibitem{Mallick} {\sc R. Mallick}, {\em Cytoplasmic Dynein Functions as a Gear in Response to Load}, Nature, 427 (2004), pp.~649--652.

\bibitem{Redner} {\sc S. Redner}, {\em A Guide to First Passage Processes}, Cambridge University Press, Cambridge, Massachussets,  2001.

\bibitem{Henrici}  {\sc P. Henrici}, {\em Applied and Computational Complex Analysis.
Vol. 3.}, John Wiley \& Sons Inc, New-York  1977.

\bibitem{Abramowitz} {\sc M. Abramowitz, and I.~A. Stegun}, {\em Handbook of Mathematical Functions}, Dover, New York 1972.
\bibitem{Carslaw} {\sc H.~S. Carslaw, and J.~C. Jaegger}, {\em Conduction of Heat in Solids}, Oxford University Press, Oxford, U.K.  1959.


\bibitem{schroer} {\sc S.~J. King and T.~A Schroer}, {\em Dynactin Increases the Processivity of the Cytoplasmic Dynein Motor}, Nat. Cell Biol., 2 (2000), pp.~20--24.

\bibitem{Seisenberger} {\sc G. Seisenberger et al.}, {\em Real-Time Single-Molecule Imaging of the Infection Pathway of an Adeno-Associated Virus}, Science, 294 (2001), pp.~1929--1932.

\bibitem{LagacheHolcman} {\sc T. Lagache et D. Holcman}, {\em Quantifying the Intermittent Transport in the Cell Cytoplasm} (submitted).

\bibitem{Sodeik1} {\sc D. Katinka, N.Claus-Henning, and B. Sodeik},
{\em Viral Stop-and-Go along Microtubules : Taking a Ride with
Dynein and Kinesins}, Trends in Microbiology, 13(7) (2005),
pp.~320--327.

\bibitem{Gross} {\sc S.~P. Gross, M.~A Welte, S.~M. Block, and E.~F. Wieschaus}, {\em Dynein-mediated Cargo Transport in Vivo : a Switch Controls Travel Distance}, The Journal of Cell Biology, 5 (2000), pp.~945--955.

\bibitem{Whittaker} {\sc G.~R. Whittaker}, {\em Virus Nuclear Import}, Advanced Drug Delivery Rewiews, 55 (2003), pp.~733--747.
\bibitem{Dean} {\sc D.~A. Dean, R.~C. Geiger, and R. Zhou}, {\em Intracellular Trafficking of Nucleic Acids}, Expert Opinion Drug Delivery, 1 (2004), pp.~127--140.
\bibitem{Campbell} {\sc E.~M. Campbell, and T.~J. Hope}, {\em Gene Therapy Progress and Prospects : Viral Trafficking During Infection}, Gene Therapy, 12 (2005), pp.~1353--1359.
\bibitem{Luo} {\sc D. Luo, and W.~M. Saltzman}, {\em Synthetic DNA Delivery Systems}, Nature Biotechnology, 18 (1999), pp.~33--37.


%




\end{thebibliography}
\end{document}